\documentstyle[11pt]{article}

\input epsf

\global\arraycolsep=1pt
\begin{document}
\def\o#1#2{{#1\over#2}}
\def\bfone{\relax{\rm 1\kern-.35em 1}}
\def\inbar{\vrule height1.5ex width.4pt depth0pt}
\def\IC{\relax\,\hbox{$\inbar\kern-.3em{\rm C}$}}
\def\ID{\relax{\rm I\kern-.18em D}}
\def\IF{\relax{\rm I\kern-.18em F}}
\def\IH{\relax{\rm I\kern-.18em H}}
\def\II{\relax{\rm I\kern-.17em I}}
\def\I1{\relax{\rm 1\kern-.28em l}}
\def\IN{\relax{\rm I\kern-.18em N}}
\def\IP{\relax{\rm I\kern-.18em P}}
\def\IQ{\relax\,\hbox{$\inbar\kern-.3em{\rm Q}$}}
\def\IZ{\relax\,\hbox{$\inbar\kern-.3em{\rm Z}$}}
\def\IR{\relax{\rm I\kern-.18em R}}
\font\cmss=cmss10 \font\cmsss=cmss10 at 7pt
\def\ZZ{\relax\ifmmode\mathchoice
{\hbox{\cmss Z\kern-.4em Z}}{\hbox{\cmss Z\kern-.4em Z}}
{\lower.9pt\hbox{\cmsss Z\kern-.4em Z}}
{\lower1.2pt\hbox{\cmsss Z\kern-.4em Z}}\else{\cmss Z\kern-.4em
Z}\fi}

\hfill    HUTP-95/A012


\hfill   March, 1995

\begin{center}
\vspace{10pt}
{\large \bf
TOPOLOGICAL FIELD THEORY
AND PHYSICS
}
\vspace{10pt}

{\sl Damiano Anselmi}

\vspace{4pt}

{\it Lyman Laboratory, Harvard University, Cambridge MA 02138, U.S.A.}

\vspace{12pt}

{\bf Abstract}
 \end{center}

\vspace{4pt}
Topological Yang-Mills theory with the
Belavin-Polyakov-Schwarz-Tyupkin $SU(2)$
instanton is solved completely, revealing an underlying
multi-link intersection theory. Link invariants are also shown to
survive the
coupling to
a certain kind of matter (hyperinstantons).
The physical relevance of
topological field theory and its
invariants is discovered.  By embedding
topological Yang-Mills theory into pure Yang-Mills theory, it is shown
that the topological version TQFT of a quantum field theory QFT
allows us to
formulate consistently the perturbative expansion of QFT in the
topologically
nontrivial sectors. In particular, TQFT
classifies the set of good measures over the instanton moduli space and
solves the inconsistency problems of the previous approaches.
The qualitatively new physical implications are pointed out.
Link numbers in QCD are related to a non abelian analogoue
of the Aharonov-Bohm effect.
\eject

\section{Introduction}
\label{intro}
\setcounter{equation}{0}

The idea that in some quantum field theories special correlation
functions are
exactly calculable dates back to the eighties,
when it was realized that supersymmetric theories possess amplitudes
(`gaugino condensates' or, in general, `topological amplitudes') that are
independent of the distances between
points\footnotemark\footnotetext{For a
review and references see \cite{rossi}. Those results are mainly
due to the
groups
Novikov-Shifman-Vainshtein-Zacharov,
Amati-Konishi-Meurice-Rossi-Veneziano,
Affleck-Dine-Seiberg.}.
A systematic study
of topological amplitudes can be pursued in the realm of the so called
topological field theories, where every physical amplitude is of this
type. Some topological field theories
can be produced by a formal procedure,
the topological twist \cite{wittenym,twist1,twist2},
starting from N=2 supersymmetric theories.
Up to now, topological field theories have not been
related in a systematic way
to their nonsupersymmetric counterparts\footnotemark
\footnotetext{Recently, nevertheless, it has been
proposed in \cite{martellini} that certain amplitudes of
topological field theory (Donaldson invariants) can be recovered in
ordinary Yang-Mills theory in a suitable limit.} and their importance for
physics has not been
identified.

According to Witten \cite{wittenym}, the physical
amplitudes of topological Yang-Mills theory are
the so-called Donaldson invariants \cite{donaldson}.
The observables ${\cal O}_{\gamma_i}$ are interpreted as cocycles of the
instanton moduli space ${\cal M}$
and are associated to cycles $\gamma_i$ of the four-manifold
$M$ ({\sl Donaldson map}).
With $M=\IR^4$ (or $S^4$), $G=SU(2)$
and unit instanton number,
there is no amplitude of this type, since
the only nontrivial cycles of  $\IR^4$ are the
point (associated to a 4-form on ${\cal M}$)
and the manifold $M$ itself: the moduli space dimension,
which is $8k-3$, $k\in \IZ$,
cannot be saturated with a product of 4-forms.
Nevertheless, in ref.\ \cite{anomali} it was explicitly
shown that the theory is not empty.
The key idea was to integrate the usual observables
of topological Yang-Mills theory on {\sl contractible} closed
submanifolds $\gamma_i$ of $\IR^4$,
rather than on cycles.
The result was that the expectation value of the product of {\sl two}
observables associated to
{\sl linked} submanifolds $\gamma_1\subset \IR^4$ and
$\gamma_2\subset \IR^4$
is indeed nonzero. The submanifolds $\gamma_1$ and $\gamma_2$ are
said to be
{\sl linked}
if $\gamma_1$ is a nontrivial cycle of $\IR^4\backslash \gamma_2$ and
$\gamma_2$ is a nontrivial cycle of $\IR^4\backslash \gamma_1$. The
idea was
tested in two cases, namely
\begin{equation}
<{\cal O}_{S^3}\cdot {\cal O}_{P}>=\backslash\!\!\!\slash
(S^3,P),\quad\quad
<{\cal O}_{S^2}\cdot {\cal O}_{S^1}>=\backslash\!\!\!\slash (S^2,S^1).
\label{ampl}
\end{equation}
The left hand side denotes the amplitudes as defined in topological
Yang-Mills
theory. The right hand side denotes the result of the explicit
computations, that was interpreted as
the {\sl link intersection number} of the $\gamma_i$'s. I use the symbol
$\backslash\!\!\!\slash$ for such a kind of intersection numbers.
$\backslash\!\!\!\slash (S^3,P)$ is a step function: zero if the
point $P$ is
placed outside the 3-sphere $S^3$; 1
if the point $P$ is placed inside $S^3$.
$\backslash\!\!\!\slash (S^2,S^1)$ is entirely similar.
In ref.\ \cite{anomali} correlation functions that do not vanish,
although
naively expected to,
were also found in four dimensional topological gravity.

The considered amplitudes involve non-local observables,
related to circles, spheres, Riemann surfaces, 1-knots, 2-knots,
etc. In N=2
super Yang-Mills theory the gaugino condensates \cite{rossi}
are average values of {\sl local} observables. It is interesting to
know that
amplidutes with nonlocal observables can also be explicitly
computed in four
dimensions
and it  would also be desirable to compute similar amplitudes
in supersymmetric theories.
Moreover, all gaugino condensates are constant amplitudes: step
amplitudes have
not been
found in supersymmetric theories, so far.

The hidden link-theory contained in topological Yang-Mills theory
deserves
to be explored in depth.
One of the purposes of this paper is to push the analysis of ref.\
\cite{anomali} forward,
answering some of the questions raised there, in particular identifying
completely the mathematical meaning of the unit instanton number
sector of the
theory.
It will be shown that this sector of  the theory contains
the full set of {\sl multilink invariants} of
closed smooth submanifolds of $\IR^4$. Multilink intersection
theory is defined
in
section \ref{multilinks}.

In a double linkage $U \backslash\!\!\!\slash V$, one usually writes
$U=\partial B$ ($B$
being called Seifert manifold) and counts the intersections $B\cap V$.
There is a well-known integral representation, due to Gauss,
of the link number between two 1-knots in $\IR^3$, namely
\begin{equation}
U \backslash\!\!\!\slash V={1\over 4\pi}
\oint_Udx^i\oint_V dy^j\varepsilon_{ijk}\partial_k {1\over |x-y|}.
\label{gauss}
\end{equation}
Topological field theory provides natural generalizations of this
formula.
The integral representation of 2-linkages in $\IR^4$ is known in the
literature, the representations of multilikages are new.

In view of the results of \cite{anomali} and the present paper, it
seems that
the generic ideas according to which N=2 supersymmetric Yang-Mills theory
in the ultraviolet is Donaldson theory and the topological twist
reorganizes the topological amplitudes of an
N=2 theory into a self-consistent topological sub-theory
deserves to be reconsidered.
Actually, the topological contents of
N=2 super Yang-Mills theory (``gaugino condensates''), topological
Yang-Mills
theory
and the so-called
``Donaldson theory'' appear to be, in general,
essentially different and only formally related by the topological twist.
Moreover, it will be suggested here that
topological Yang-Mills theory is more intrinsically related to ordinary
Yang-Mills theory ({\sl via} a certain {\sl topological embedding}, see
section \ref{embedding})
than to N=2 supersymmetric Yang-Mills theory.
Indeed, it is one of the main purposes of this paper to propose and
study the
role
of topological field theory in physics. The topological amplitudes
are expected to carry some physical information. Link numbers in
QCD should be
related to a
non abelian analogue
of the Aharonov-Bohm effect (which could be  in principle detectable).

The organization of the paper is the following. In section
\ref{multilinks},
multilink intersection theory is defined and tested by computing various
amplitudes and by working out integral representations of multilink
invariants.
In section \ref{coupling} the so-called hyperinstantons, introduced
and studied
by Fr\`e and the author in ref.s \cite{twist2,hyper,ghyp}, are used
to show
that the properties of pure topological
Yang-Mills theory survive the coupling to matter (scalar fields, in
this case).
In section \ref{embedding}, the topological embedding is realized
and compared
to the
usual treatment of collective coordinates. Finally,
in section \ref{aharonov}
the relation between link numbers
and Aharonov-Bohm effect is discussed.

\section{Multilinks}
\label{multilinks}
\setcounter{equation}{0}

The aim of this section is to define multilink intersection theory,
work out
its relation with topological Yang-Mills theory,  test this
relation and find
integral representations of multilink invariants.

According to the common interpretation, topological field theory
deals  with
intersection theory on the moduli space ${\cal M}$ of some instantons
on a manifold $M$. This means that the average value
of a product of observables ${\cal O}_{\gamma_i}$ integrated over
cycles $\gamma_i\subset M$ has an interpretation
\begin{equation}
<{\cal O}_{\gamma_1}\cdots {\cal O}_{\gamma_n}>=\# (L_1,\cdots ,L_n)
\label{2.1}
\end{equation}
as {\sl intersection number} (here denoted with $\#$) of cycles
$L_i\subset
{\cal M}$
associated to the $M$-cycles $\gamma_i$.
The above expression is well-defined and possibly
nonzero only when the intersection on the right hand side
is a {\sl complete} intersection, which means
\begin{equation}
\sum_{i=1}^n {\rm codim}\, L_i={\rm dim}\, {\cal M}.
\label{con}
\end{equation}
In that case, the intersection $\cap_i L_i$ is a discrete set of
points and the
operation $\#$
{\sl counts} these points (with a suitable sign assignement that
can be defined
rigorously).

Now, the results of ref.\ \cite{anomali}
imply that this is not the whole story about topological
field theories in four dimensions, in general.
Indeed, link numbers belong to a quite different class of invariants.
Nevertheless, one expects that the interpretation of  these new
topological
correlation functions is formally similar to the above one,  once
one replaces
the symbol $\#$ with $\backslash\!\!\!\slash$ and
$L_i$ with $\gamma_i$ themselves. Moreover,
$\gamma_i$ are
closed $M$-submanifolds, but not necessarily $M$-cycles. Thus, we
expect to
have,
instead of (\ref{2.1}),
\begin{equation}
<{\cal O}_{\gamma_1}\cdots {\cal O}_{\gamma_n}>=\backslash\!\!\!\slash
(\gamma_1,\cdots ,\gamma_n),
\label{amplex}
\end{equation}
and that this expression is possibly nontrivial when some analogue of
(\ref{con}) holds.

For simplicity, I assume that none of the $\gamma_i$ is $M$ itself.
In general, the $\gamma_i$'s wil be compact. However, in the explicit
calculations it
is sometimes convenient to `uncompactify' them, for example describing
a 2-sphere as a 2-plane.

The above symbolic expressions suggest that one should be able define a
suitable concept
of multi-linkage and a suitable criterium of {\sl complete} multilinkage.

\phantom{.}

\underline{\sl Completeness}.

A multiple intersection point $P$
among a set of $M$-submanifolds
$\gamma_1,\ldots \gamma_n$ is {\sl complete} if
\begin{equation}
\sum_{i=1}^n {\rm codim}\, \gamma_i={\rm dim}\, M+1.
\label{compl}
\end{equation}
$P$ is called a {\sl complete intersection}.
$\backslash\!\!\!\slash (\gamma_1,\ldots \gamma_n)$ is called a
{\sl complete multilink intersection form}. All situations in which
(\ref{compl}) does not hold are referred to as {\sl incomplete
intersections}.

In relation (\ref{compl})
(as well as in the multilink problem) there is no trace either of $SU(2)$
instantons,
or topological Yang-Mills amplitudes.
The funny fact is that for $M=\IR^4$ (or $S^4$), ${\rm dim}\, M+1$ equals
the dimension of the moduli space of $SU(2)$ instantons on $M$ with unit
instanton number.
Condition (\ref{compl})  is clearly satisfied by the amplitudes
(\ref{ampl}).
Indeed, from the field theoretical point of view, (\ref{compl}) is
nothing but
the requirement that the ghost number anomaly should be compensated
by the sum
of the ghost numbers of the observables.

For $n=2$, (\ref{compl}) can be written in the form
${\rm dim}\, \gamma_1+{\rm dim}\, \gamma_2={\rm dim}\, M-1$,
which is the usual rule for 2-linkages. However,
this form is not suitable for the multilink generalization:
the correct expression is (\ref{compl}).

\phantom{.}

\underline{\sl Multi-linkage}.

The idea of multilinkage is the following. Consider the amplitude
(\ref{amplex}). One wants
to deform the $\gamma_i$'s smoothly in $\IR^4$ in orther to {\sl
unlink} them
or contract them to points: in practice,
to move them very far from one another. In doing this, two things
can happen:

i) a {\sl proper} subset of the $\gamma_i$'s intersect in some
point or some
locus of points.
This is an incomplete intersection and is valued $0$.
The reason for this is that, when the $\gamma_i$'s satisfy (\ref{compl}),
then no proper subset of the $\gamma_i$'s can satisfy an
analogous relation
and intersect {\sl completely}.
For example, two 2-spheres $S^2$ and ${S^2}^\prime$ can be
intersected and
superposed without problems.

ii) in the movement, it is necessary to cross complete intersections,
i.e.\ points in which {\sl all} the
$\gamma_i$'s intersect contemporarily.
Each of these points contributes with one unit.
The multilink intersection number is the (algebraic) counting of
these points.
It is easy to see that, in a generic situation, this is the counting of a
discrete number of  points.

The rigorous definition of the signs of each contribution is
encoded in the
explicit integral formul\ae\ that will be derived from topological field
theory.

\phantom{.}

The following property holds:
\begin{equation}
\backslash\!\!\!\slash (\gamma_1,\ldots \gamma_i,\gamma_{i+1},\ldots
\gamma_n)=(-1)^{{\rm codim}\, \gamma_i \cdot {\rm codim}\,
\gamma_{i+1}}\,\,
\backslash\!\!\!\slash (\gamma_1,\ldots \gamma_{i+1},\gamma_i,\ldots
\gamma_n),
\end{equation}
that means that one has to take the order in which the $\gamma_i$'s
are listed
into account.

Considering the example
$\hbox{$\backslash\!\!\!\slash(S^3,{S^3}^\prime,S^1)$}$
of a triple linkage with two 3-spheres $S^3$ and
${S^3}^\prime$ and one circle $S^1$, depicted in the figure,
one can easily work out alternative definitions of the multilink
intersection
points.
The picture is at fixed time $t=0$: the 3-spheres appear as
2-spheres at fixed
time. The circle $S^1$ is at $t\equiv 0$. $S^1$ intersects $S^3$ in
the points
$P$ and $S$
and ${S^{3}}^\prime$ in $R$ and $Q$.
The intersection among
the two 3-spheres is a 2-sphere $S^2$.
At $t=0$ such an intersection appears as a circle $C$.
$C$ is the boundary of a surface $D$ (the shadowed region in the
picture),
that intersects $S^1$ in a point $T$.
The counting of the points $T$ obtained in this way with
appropriate signs gives the amplitude.
Describing the set of points that have to be counted does not
seem so difficult, but it seems nontrivial to assign
appropriate signs to them. Topological Yang-Mills theory already
contains the correct prescription.

\let\picnaturalsize=N
\def\picsize{5.0in}
\def\picfilename{link.eps}
\ifx\nopictures Y\else{\ifx\epsfloaded Y\else\fi
\global\let\epsfloaded=Y
\centerline{\ifx\picnaturalsize N\epsfxsize \picsize\fi
\epsfbox{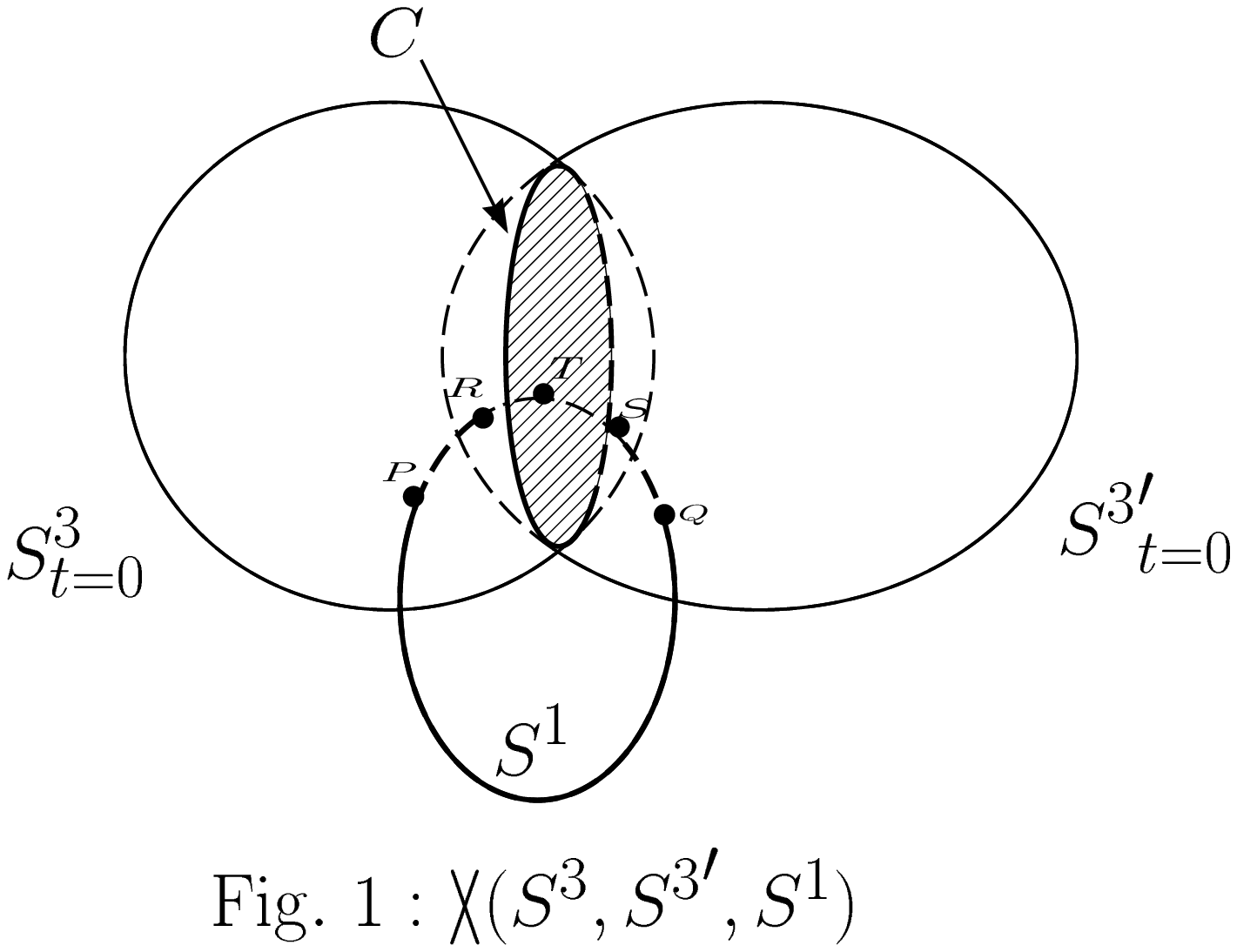}}}\fi

To show that the above construction is meaningful, one has to prove that
it does not depend on the chosen couple of $\gamma_i$'s one starts with.
So, let us consider, now, the following
alternative possibility. This time, we start by
considering the intersection between $S^3$ and $S^1$,
which is represented by the two points $P$ and $S$.
$P$ and $S$ are the boundary of a segment
$\overline{PS}$. $\overline{PS}$ meets ${S^3}^\prime$
in a point $N$ (not shown in the picture).
The amplitude can also be described as the counting
of such points.

A third equivalent description is the following. Let $\beta_1$ be
such that $\partial\beta_1=\gamma_1$. Then, consider the
intersection among
$\beta_1$
and the other $\gamma_i$'s, $i=2,\ldots n$. This is in general a
discrete set
of points and  counting them gives the amplitude.

\phantom{.}

\underline{\sl Proposition.}
Multilink intersection theory is the solution to topological Yang-Mills
theory with $G=SU(2)$, $M=\IR^4$ and unit instanton number.

\phantom{.}

I am now going to check the above proposition in various cases of
multilink
intersections
and derive their integral representations.
It is useful to report here the explicit solution of the theory,
as it was elaborated in ref.\ \cite{anomali}, to which the reader
is referred
for
the details of the derivation (see also formula (\ref{por})).
The solution is encoded into the generating expressions
\begin{eqnarray}
\hat {\cal Q}&=&{1\over 16 \pi^2}\hat F^a\hat F^a={1\over 4\pi^2}
{\rho^3\over D^4}[\rho dV(x-x_0)-4d\rho \wedge (x-x_0)^\mu
d\sigma_\mu(x-x_0)]={1\over 16 \pi^2}\hat d \hat C,\nonumber\\
\hat C&=&\hat A^a\hat F^a-{1\over 6}\varepsilon_{abc}\hat A^a
\hat A^b \hat A^c={4\over 3}{1\over D^3}[3\rho^2+(x-x_0)^2]
(x-x_0)^\mu d\sigma_\mu(x-x_0).
\label{hal}
\end{eqnarray}
$(\rho, x_0)\in {\cal M}=(0,\infty)\otimes \IR^4$ are
the five
moduli, while $x\in M=\IR^4$ and  $D=\rho^2+(x-x_0)^2$.
$F^a=dA^a+{1\over
2}{\varepsilon^a}_{bc}A^b A^c$ denotes the field strength. $\hat
F^a=F^a+\psi_0^a+\phi_0^a$
and $\hat A^a=A^a+C^a_0$ are the relevant BRST extensions. Finally,
$dV(x)=dx^\mu d\sigma_\mu(x)$,
$d\sigma_\mu(x)=\varepsilon_{\mu\nu\rho\sigma}
dx^\nu dx^\rho dx^\sigma$ and $\hat d=d+s$, $s$ being the BRST operator
\begin{equation}
s=dm^i{\partial\over \partial m^i}=d \rho{\partial\over \partial \rho}+d
x_0^\mu{\partial\over \partial x_0^\mu}.
\label{brstop}
\end{equation}
The observables are
\begin{equation}
{\cal O}_\gamma^{(n)}=\int_{\gamma}\hat {\cal Q}^n.
\label{obser}
\end{equation}

It was noticed in \cite{anomali}
that the amplitudes are ``automatically'' normalized correctly by
the factor
${1\over 16\pi^2}$ appearing in (\ref{hal}).
This property gives an intrinsic and concrete
meaning to formal concepts and set-ups like
the `BRST extension' (see \cite{baulieu,twist1,twist2,hyper}).
The notation of ref.\ \cite{anomali} is strictly followed in the
sequel, with
the only difference
that the observables ${\cal O}$ and the corresponding
${\cal M}$-differential forms
are defined {\sl ab-initio} with the correct normalization factor
${1\over 16\pi^2}$.

Any topological amplitude can be written as an integral over the
boundary of
the moduli space
${\cal M}$. For $SU(2)$ instantons on $S^4$ such a boundary
corresponds to
$\rho=0$, while for instantons on $\IR^4$ the boundary possesses more
components
($\rho\rightarrow \infty$ and $x_0\rightarrow \infty$).
Nevertheless, these extra components never contribute to the amplitudes
computed in
ref.\ \cite{anomali}, confirming that  one can safely extend the
results to
topological Yang-Mills theory on $S^4$. In this case, however, one should
imagine that a puncture is placed at infinity.
In the appendix to this section it is proved on general grounds, that
the $\partial {\cal M}$-component
$x_0\rightarrow \infty$ never contributes.
Knowing this a priori is useful to simplify the computations, since
in many cases, it is convenient to `uncompactify' some submanifolds
$\gamma_i$
and
it is not correct to check the $x_0\rightarrow \infty$
component {\sl after} having uncompactified.
I shall comment on the vanishing of the
$\rho\rightarrow \infty$ terms along with the computations.

The $\rho\rightarrow 0$ limit of (\ref{hal}) can be done easily,
using the
property
\begin{equation}
\lim_{\rho\rightarrow 0}{\rho^4\over (\rho^2+x^2)^4}={\pi^2\over 6}
\delta(x),
\label{using}
\end{equation}
so that on $\partial {\cal M}$ ($d\rho$ being also zero)
\begin{equation}
\hat{\cal Q}(x)\rightarrow{1\over 4!}\delta(x-x_0)dV(x-x_0)=
-{1\over  4!\pi^2} \hat d\,
\partial_\mu{1\over (x-x_0)^2} \,d\sigma_{\mu}(x-x_0).
\label{delta}
\end{equation}
In practice, everything is encoded into these very simple
expressions, although
it is (\ref{hal})
that makes any computation meaningful. Some properties, that are
not visible
from
(\ref{delta}) will follow immediately from (\ref{hal}). Check, for
example, the
end of subsection \ref{green}.

Now, we are ready to begin the computations.
First (Example 1) the integral representation of
$\hbox{$\backslash\!\!\!\slash(\gamma_1,\gamma_2)$}$ is derived from
topological field theory.
Then, we proceed by studying genuine multilink invariants and
extracting their
integral representations.
The calculation of the first multilink intersection number (Example
2) is done
in full detail,
the other computations being simply scketched.

\subsection{Evaluations}
\label{evalua}

\hbox{\phantom{sing}\underline{Example 1.} Integral representation of
$\hbox{$\backslash\!\!\!\slash(\gamma_1,\gamma_2)$}$.}

Let us consider two submanifolds $\gamma_1,\gamma_2\subset \IR^4$, of
dimensions $1$ and $2$, respectively. Let the corresponding
observables be
${\cal O}_{\gamma_1}=\omega^{(3)}_{\gamma_1}=d\Omega^{(2)}_{\gamma_1}$
and ${\cal
O}_{\gamma_2}=\omega^{(2)}_{\gamma_2}=d\Omega^{(1)}_{\gamma_2}$.
We have
\begin{equation}
\hbox{$\backslash\!\!\!\slash(\gamma_1,\gamma_2)$}=
<{\cal O}_{\gamma_1}\cdot {\cal O}_{\gamma_2}>
=\int_{{\cal M}}\omega^{(3)}_{\gamma_1}\wedge \omega^{(2)}_{\gamma_2}=
\int_{\partial {\cal M}}\Omega^{(2)}_{\gamma_1}
\wedge \omega^{(2)}_{\gamma_2}.
\end{equation}
$\Omega^{(2)}_{\gamma_1}$ and
$\omega^{(2)}_{\gamma_2}$ can be easily written down from
(\ref{hal}). We have
thus
\begin{equation}
\backslash\!\!\!\slash(\gamma_1,\gamma_2)=- {3\over 2\pi^4}
\int_{\gamma_1}dx^\mu\int_{\gamma_2}dy^\nu dy^\rho
\lim_{\rho\rightarrow 0}\int_{\IR^4}d^4x_0
{\varepsilon_{\mu\nu\rho\sigma}
(x-x_0)^\sigma  \rho^4[3\rho^2+(x-x_0)^2]\over [\rho^2+(x-x_0)^2]^3
[\rho^2+(y-x_0)^2]^4}.
\end{equation}
It is convenient to write
\begin{equation}
\backslash\!\!\!\slash(\gamma_1,\gamma_2)=
\int_{\gamma_1}dx^\mu\int_{\gamma_2}dy^\nu dy^\rho
\varepsilon_{\mu\nu\rho\sigma}
V^\sigma(x-y),
\end{equation}
where
\begin{equation}
V^\sigma(z)=
-{3\over 2\pi^4}\lim_{\rho\rightarrow 0}
\int_{\IR^4}d^4x_0
{(z-x_0)^\sigma \rho^4 [3\rho^2+(z-x_0)^2]\over
[\rho^2+(z-x_0)^2]^3 (\rho^2+x_0^2)^4}=
-{1\over 4\pi^2}{z^\sigma\over |z|^4}.
\end{equation}
The form of the last expression follows from dimensional considerations.
The constant $-{1\over4\pi^2}$ is determined after rescaling
all quantities by $\rho$ and applying the dominated convergence theorem,
or, more quickly, but less rigorously, by using (\ref{using}).
We conclude
\begin{equation}
\hbox{$\backslash\!\!\!\slash(\gamma_1,\gamma_2)$}=
{1\over 8\pi^2}\int_{\gamma_1}dx^\mu\int_{\gamma_2}dy^\nu dy^\rho
\varepsilon_{\mu\nu\rho\sigma}\partial_\sigma {1\over (x-y)^2},
\label{inte}
\end{equation}
which is the desired expression. Notice the appearance of the Green
function
$ {1\over (x-y)^2}$, which is singular precisely on the complete
intersection
points $x=y$.
Of course, (\ref{inte}) contains (\ref{gauss}): writing $\IR^4$ as
$\IR^3\otimes \IR$ and
taking $U=\gamma_1\subset \IR^3$, $\gamma_2=V\otimes \IR$,
$V\subset \IR^3$,
(\ref{inte}) gives back (\ref{gauss}).

\underline{Example 2.}
$\hbox{$\backslash\!\!\!\slash(S^3,{S^3}^\prime,S^1)$}$.

In the first example of multilinkage, I
consider a triple intersection
among two 3-spheres and one circle (see also Fig.\ 1). The
3-spheres will be
`uncompactified'
to $\IR^3$'s, this simplifying a bit the calculation.
Precisely, I choose
\begin{equation}
S^3=\IR^2\times \IR \times \{0\},\quad\quad
{S^3}^\prime=\IR^2\times \{\bar x\} \times \IR,\quad\quad
S_r^1=\{(0,0)\}\times C_r,
\end{equation}
$C_r$ denoting a circle of radius $r$ placed in the last $\IR^2$ and
centered in the origin. The above submanifolds of $\IR^4$
are indeed multilinked according
to the general definition. $S^3$ and ${S^{3}}^\prime$ intersect in the
`time' axis (which, by convention, is the first one, while the other axis
are $x$, $y$ and $z$, in the order) and in the $x$-axis.
Such intersections are irrelevant, because incomplete.
The circle $S^1$ is at fixed time $t=0$ and winds around
the $x$-axis.
It intersects $S^3$ in a couple of points on the $y$-axis
and ${S^3}^\prime$ in a couple of points on the $z$-axis.
These intersections are also incomplete.
Now, if $r>\bar x$,
when one wants to {\sl unlink} the three objects,
it is necessary to cross precisely one complete (triple)
intersection point.
Thus, we predict that the corresponding topological amplitude is
equal to one.
Instead, if $r<\bar x$, the unlinking procedure goes on safely
and we predict a zero amplitude.

I want to compute
\begin{equation}
{\cal A}=<{\cal O}_{S^3} \cdot {\cal O}_{{S^3}^\prime}\cdot {\cal
O}_{S^1_r}>=
\int_{\cal M}\omega^{(1)}\wedge {\omega^{(1)}}^\prime
\wedge \omega^{(3)}_r.
\end{equation}
Using the first of
(\ref{hal}), it is easy to verify that the ${\cal M}$-1-forms
corresponding to the first two observables are
\begin{equation}
\omega^{(1)}=\int_{S^3}\hat Q={3\over 4}
{\rho^3 [\rho dx_0^4-x_0^4 d\rho]\over [\rho^2+
(x_0^4)^2]^{5\over 2}},\quad\quad
{\omega^{(1)}}^\prime=\int_{{S^3}^\prime}\hat Q={3\over 4}
{\rho^3 [\rho dx_0^3-(x_0^3) d\rho]\over [\rho^2+
(x_0^3-\bar x)^2]^{5\over 2}}.
\label{uno}
\end{equation}
An immediate check shows that these differential forms are closed,
as it must be.

Instead, in order to write down $\omega^{(3)}_r$, it is convenient
to use the second of (\ref{hal}), to express $\omega^{(3)}_r$
as $d\Omega^{(2)}_r$, and ${\cal A}$ as
$\int_{\partial {\cal M}}\omega^{(1)}\wedge {\omega^{(1)}}^\prime
\wedge \Omega^{(2)}_r$. One finds
\begin{equation}
\Omega^{(2)}_r={1\over 4\pi^2}\int_{C_r}{3\rho^2+(x_0^1)^2+(x_0^2)^2+
({\bf x}-{\bf x}_0)^2\over [\rho^2+(x_0^1)^2+(x_0^2)^2+
({\bf x}-{\bf x}_0)^2]^3}(x-x_0)^\mu dx^\nu\varepsilon_{\mu\nu\rho\sigma}
dx_0^\rho dx_0^\sigma,
\end{equation}
the bold-face denoting 2-vectors in the plane generated by the $y$-axis
and the $z$-axis.

We know that the only  component of the boundary $\partial {\cal M}$
that contributes is the one for $\rho\rightarrow 0$.
So, we can put $d\rho=0$ and take the limit $\rho\rightarrow 0$.
Let us also choose $\bar x=0$, for now. We have
\begin{equation}
{\cal A}=\lim_{\rho\rightarrow 0}{9\rho^8\over 32 \pi^2}
\int_{\IR^4}d^4x_0\int_0^{2\pi}d\theta
{r(r-x_4 \cos\theta)
(3\rho^2+x_0^2+r^2-2 r x_4 \cos\theta)\over
[\rho^2+(x_0^4)^2]^{5\over 2}
[\rho^2+(x_0^3)^2]^{5\over 2}
(\rho^2+
x_0^2+r^2-2 r x_4 \cos \theta)^3},
\end{equation}
where $x_4=\sqrt{(x_0^3)^2+(x_0^4)^2}$. Now, rescaling all quantities
by $\rho$, redefining $r$ as $r\over \rho$ and
integrating over $x_0^1$, $x_0^2$ and $\theta$
one easily arrives at
\begin{equation}
{\cal A}=\lim_{r\rightarrow \infty}{9\over 32 \pi}
\int_{-\infty}^\infty dx_0^4
\int_{-\infty}^\infty dx_0^3
{\varphi(x_4,r)\over
[1+(x_0^4)^2]^{5\over 2}
[1+(x_0^3)^2]^{5\over 2}},
\end{equation}
where
\begin{equation}
\varphi(x_4,r)=\pi+\pi{(r^2-x_4^2)^3+3(r^4-x_4^4-x_4^2)+r^2-1
\over [(r^2-x_4^2)^2+ 2(r^2+x_4^2)+1]^{3\over 2}}.
\end{equation}
The function $\varphi(x_4,r)$ tends to zero for $x_4\rightarrow \infty$,
to $2\pi$ for $r\rightarrow \infty$, is continuous and bounded.
So, by the dominated convergence theorem, we can exchange the
limit and the integration, finally obtaining
\begin{equation}
{\cal A}={9\over 16}
\left(\int_{-\infty}^\infty {dy\over (1+y^2)^{5\over 2}}\right)^2=1,
\end{equation}
confirming the expectation.
Moreover, $\varphi(x_4,0)\equiv 0$.
So, for $r\rightarrow 0$, which corresponds to $\rho\rightarrow\infty$,
we have a zero result, confirming that
the $\partial {\cal M}$-component $\rho\rightarrow\infty$ does
not contribute.

Taking $\bar x\neq 0$, there are
{\sl two} quantities that, after the rescaling by $\rho$,
should tend to infinity: the rescaled
$r$ and the rescaled $\bar x$. The order according to which these limits
should to be taken is dictated by which one of the inequalities
$r>\bar x$ and $r<\bar x$ is true. In the first case, everything
goes on as
before,
but in the second case one gets zero. We conclude
\begin{equation}
{\cal A}(r,\bar x)={1\over 2}(1+H(r-\bar x)).
\end{equation}
where $H(x)=1$ for $x>0$ and $H(x)=-1$ for $x<0$:
the amplitude is a step function. Finally, one can check that for
$r=\bar x$
the result is ${1\over 2}$, so that we can define $H(0)=0$.

\underline{Example 3.}
$\backslash\!\!\!\slash(S^3_{(1)},S^3_{(2)},S^3_{(3)},S^2_r)$.

In this example, the 3-spheres will be described by $\IR^3$'s, at $x=0$,
$y=0$ and $z=0$, respectively.
The corresponding ${\cal M}$-forms are the analogues of (\ref{uno}).
Thus, we have a triple intersection, which is the entire time axis at
$x=y=z=0$.
The 2-sphere $S^2_r$ will be placed at fixed time $t=0$ and
centered in the
origin of the
3-space generated by the $x$, $y$ and $z$-axis. Vectors in such a
three space
will be written in boldface. It is obvious that one cannot unlink
the four
objects without meeting one complete
intersection point.
Using the second of (\ref{hal}) we write ${\cal A}=\int_{\partial {\cal
M}}\prod_{i=1}^3\omega^{(1)}_i\Omega^{(1)}_r$, with
$\omega^{(2)}_r=d\Omega^{(1)}_r$
and
\begin{equation}
\Omega^{(1)}_r={1\over 4\pi^2}\int_{S^2_r}{[3\rho^2+(x_0^1)^2+({\bf
x}-{\bf
x}_0)^2]\,
(x-x_0)^\mu\varepsilon_{\mu\nu\rho\sigma}dx^\nu dx^\rho dx_0^\sigma\over
[\rho^2+(x_0^1)^2+({\bf x}-{\bf x}_0)^2]^3}.
\end{equation}
Thus, after rescaling all quantities by $\rho$ and integrating ove
$x_0^1$,
one easily arrives at
\begin{equation}
{\cal A}=\lim_{r\rightarrow \infty}{3^3\over 2\cdot 4^3}\int
\prod_{i=2}^4dx_0^i
\int_0^\pi d\theta
{r^2 \sin\theta (r-|{\bf x}_0| \cos \theta)\over
\prod_{i=2}^4 [1+(x_0^i)^2]^{5\over 2} (r^2+|{\bf x}_0|^2-2 r |{\bf
x}_0| \cos
\theta)^{3\over 2}}.
\end{equation}
Finally, taking the limit $r\rightarrow \infty$, one finds
\begin{equation}
{\cal A}={3^3\over 4^3}
\left(\int_{-\infty}^\infty {dy\over (1+y^2)^{5\over 2}}\right)^3=1.
\end{equation}

\underline{Example 4.}
$\backslash\!\!\!\slash(S^3_{(1)},S^3_{(2)},S^3_{(3)},S^3_{(4)},S^3_{(5)})$.

Now we want to check the predictions with a 5-tuple linkage
among 3-spheres. The first four 3-spheres
$S^3_{(i)}$ $i=1,\ldots 4$ will be in fact $\IR^3$'s,
at $t=\bar t$, $x=\bar x$, $y=\bar y$ and $z=\bar z$, respectively.
They intersect in the point $\bar x=(\bar t ,\bar x,\bar y, \bar
z)\in\IR^4$.
This is a 4-intersection and so
does not contribute, according to the general rules: in the
case at hand a complete intersection is a 5-intersection.
The fifth 3-sphere $S^3_{(5)}$ (which will be really a compact sphere,
the radius being $r$) is chosen to surround the origin.
Consequently, if $r>|\bar x|$ the unlinking process necessarily
meets a 5-tuple
intersection
point and the amplitude
${\cal A}=\int_{\cal M}\prod_{i=1}^5\omega^{(1)}_i$
is expected to be equal to one. If, instead,
$r<\bar x$, the unlinking process finds no obstacle and the result
is zero.

The expressions of $\omega^{(1)}_i$, $i=1,\ldots 4$ are easily
read from (\ref{uno}), while, using some other
results of \cite{anomali}, see formul\ae\ (4.21) and (4.24) there,
one can write $\omega^{(1)}_5=df$,
where $f$ is a function that tends to ${1\over 2}(1+H(r-x_0))$
for $\rho\rightarrow 0$. Consequently, we have
\begin{equation}
{\cal A}=\lim_{\rho\rightarrow 0}{3^4\over 4^4}\int_{\IR^4}
{1\over 2}(1+H(r-|x_0+\bar x|))
{\rho^{16}\,d^4x_0\over \prod_{i=1}^4 [\rho^2+(x_0^i)^2]^{5\over 2}}.
\end{equation}
Rescaling every quantity by $\rho$ as usual,
one gets
\begin{equation}
{\cal A}={3^4\over 4^4}\left(\int_{-\infty}^\infty
{dy\over (1+y^2)^{5\over 2}}\right)^4
{1\over 2}(1+H(r-\bar x))={1\over 2}(1+H(r-\bar x)),
\end{equation}
as desired.

\subsection{Integral representations of multilink invariants}
\label{green}

The purpose of this subsection is to extract the integral
representations of
multilink invariants provided by the instanton. I focus on
$\backslash\!\!\!\slash(\gamma_3,\gamma_3^\prime,\gamma_1)$.
It is easy to see that the integral representation of
$\backslash\!\!\!\slash(\gamma_3,\gamma_3^\prime,\gamma_1)=
\int_{{\cal M}}
\omega^{(1)}_{\gamma_3}\wedge\omega^{(1)}_{\gamma_3^\prime}\wedge
\omega^{(3)}_{\gamma_1}=
\int_{\partial {\cal M}}
\omega^{(1)}_{\gamma_3}\wedge\omega^{(1)}_{\gamma_3^\prime}\wedge
\Omega^{(2)}_{\gamma_1}$ provided by the instanton can be written as
\begin{equation}
\backslash\!\!\!\slash(\gamma_3,\gamma_3^\prime,\gamma_1)=
\int_{\gamma_3}\int_{\gamma_3^\prime}\int_{\gamma_1}
d\sigma(x)\cdot dy\,\, d\sigma(x^\prime)\cdot V(x,x^\prime,y)-
d\sigma(x^\prime)\cdot dy\,\, d\sigma(x)\cdot V(x,x^\prime,y),
\end{equation}
where $V^\mu(x,x^\prime,y)={\partial\over \partial
y^\mu}v(x,x^\prime,y)$.
Here, $v(x,x^\prime,y)$ is a kind of  ``three-body Green function''. Its
expression is
\begin{equation}
v(x,x^\prime,y)={1\over 4\pi^6}\lim_{\rho\rightarrow 0}\rho^8\int_{\IR^4}
d^4x_0{1\over [\rho^2+(y-x_0)^2][\rho^2+(x-x_0)^2]^4
[\rho^2+(x^\prime-x_0)^2]^4}.
\end{equation}
 $v(x,x^\prime,y)$ should be regarded as a distribution and
can be easily determined using $(\ref{using})$. Alternatively,
it is convenient to study $\Box_y v(x,x^\prime,y)\equiv
v_y(x,x^\prime,y)$,
picking up a test function $\varphi(z,t)$, $z=x-y$, $t=x^\prime-y$,
and acting on it with $v_y$. We have
\begin{eqnarray}
v_y(\varphi)&=&\int v_y(x,x^\prime,y)\varphi(z,t)dz \, dt
\nonumber\\
&=&-{2\over \pi^6}\lim_{\rho\rightarrow 0}\int
{\rho^{10}\varphi(z,t)\, dz \,
dt\,
dx_0\over (\rho^2+x_0^2)^3[\rho^2+(z-x_0)^2]^4 [\rho^2+(t-x_0)^2]^4}
=-{1\over 36}\varphi(0,0),
\end{eqnarray}
after rescaling $z$, $t$ and $x_0$ by $\rho$, as usual.
We conclude
\begin{equation}
v_y(x,x^\prime,y)=-{1\over 36}\delta(x-y)\delta(x^\prime-y),\quad\quad
v(x,x^\prime,y)={1\over 36\pi^2}{\delta(x-x^\prime)\over
(x-2y+x^\prime)^2}.
\label{solui}
\end{equation}
The final expression of the triple link number is thus
\begin{equation}
\backslash\!\!\!\slash(\gamma_3,\gamma_3^\prime,\gamma_1)=
{1\over
36\pi^2}\int_{\gamma_3}d\sigma^\mu(x)\int_{\gamma_3^\prime}d\sigma^\nu(x^\prime)
\int_{\gamma_1}dy^\rho\,\delta(x-x^\prime)\,{\partial\over \partial
y^\sigma}{(\delta_{\mu\rho}\delta_{\nu\sigma}-\delta_{\mu\sigma}\delta_{\nu\rho})\over
(x-2y+x^\prime)^2}.
\label{soluti}
\end{equation}
This expression, as well as (\ref{solui}), is in complete agreement
with the
multilink idea.
The delta function projects onto the instersection of two
submanifolds, the
rest counts the links with the third manifold. Notice that
the Green function $v(x,x^\prime,y)$ is not symmetric in $x$,
$x^\prime$ and
$y$, but
keeps trace of the choice of the two submanifolds that are intersected.
This choice corresponds to the choice of which differential form is
converted
from $\omega^{(n)}_\gamma$ to $\Omega^{(n-1)}_\gamma$ (where
$\omega^{(n)}_\gamma=d\Omega^{(n-1)}_\gamma$)
when passing from the integral over ${\cal M}$ to the integral over the
boundary $\partial {\cal M}$. So, the instanton provides an easy
way to prove
that the result is independent of the choice
of the intersected submanidolds, although this independence is not
apparent in
formula (\ref{soluti}). The above expression was found converting
$\omega^{(3)}_{\gamma_1}$ to $\Omega^{(2)}_{\gamma_1}$.
Alternatively, doing
the same work with $\gamma_3$, instead of $\gamma_1$, one finds the
equivalent
integral representation
\begin{equation}
\backslash\!\!\!\slash(\gamma_3,\gamma_3^\prime,\gamma_1)=
{1\over
36\pi^2}\int_{\gamma_3}d\sigma^\mu(x)\int_{\gamma_3^\prime}d\sigma^\nu(x^\prime)
\int_{\gamma_1}dy^\nu\,\delta(y-x^\prime)\,{\partial\over \partial
x^\mu}{1\over
(y-2x+x^\prime)^2}.
\label{solutin}
\end{equation}
In all the other cases, Green functions and integral
representations can be
worked out similarly.

\subsection{Open problems}

We have estabilished that
multilink intersection theory  is the solution to topological
Yang-Mills theory
on $\IR^4$ (or $S^4$) with $G=SU(2)$ and unit instanton number.
Open problems concern the interpretation (and calculation) of
nonvanishing amplitudes for the other instanton numbers
\cite{rebbi,atiyah}, as well as for other gauge groups $G$ and
manifolds $M$.
For
$G=SU(N)$ and unit instanton number, the amplitudes are the same,
since the
instanton is the same. When $G=SU(3)$ the formal dimension of the
moduli space
is $12k-8$,
while for $SU(2)$ it is $8k-3$.
For $k=1$, one has $4$ instead of $5$. This is because, embedding
the $SU(2)$
instanton
in the first three generators of $SU(3)$, there is room for an
antighost zero
mode
(the constant, which is indeed meaningful on $S^4$, but not on $\IR^4$)
associated to the eight $SU(3)$ generator.
A reasonable way to define amplitudes (denoted with $\ll\ldots\gg$) with
selection rule $4$
instead of $5$ can be
\begin{equation}
\ll {\cal O}_1\cdots {\cal O}_n\gg=\lim_{r\rightarrow \infty}<{\cal
O}_{S^3_r}\cdot
{\cal O}_1\cdots {\cal O}_n>.
\end{equation}
In this way, one has, for example, using the first of (\ref{ampl}),
\begin{equation}
\ll{\rm tr}[\phi^2](x)\gg=\lim_{r\rightarrow \infty}<{\cal
O}_{S^3_r}\cdot
{\rm tr}[\phi^2](x)>=1,
\label{ko}
\end{equation}
since the infinitely large 3-sphere necessarily contains the point $x$.
Formula (\ref{ko}) is an example of a recursion relation between
invariants
for different $N$'s, perhaps a particular case of a much richer set of
recursion relations.

As far as higher instanton numbers are concerned and $G=SU(2)$, we
can make the
following comments. For
generic $k$ a very simple amplitude is $<{\cal O}_{S^3_r}\cdot
\prod_{i=1}^{2k-1}
{\rm tr}[\phi^2](x_i)>$ \cite{anomali}. A possible meaning of this
amplitude
is that it counts the number of points $x_i$ that are placed inside
$S^3_r$.
Apart from this very simple case, however, it is not so easy to identify
the meaning of the more complicated amplitudes.
Take, for example, $k=3$ and $<\prod_{i=1}^3 \int_{\gamma_i}{\cal
Q}^{2}>$,
$\gamma_i$, $i=1,2,3$, being three circles in $\IR^4$.
This should be something like the generalized link number of three
circles
$\gamma_i$.
If this has a meaning, the meaning should be nontrivial. One can
expect that
there is a canonical way of associating a 2-sphere or, in general,
a 2-knot to a couple of circles and that the amplitude is the
linking number
between the 2-knot and the third circle. Moreover, this result shoul be
independent of the choice of the initial couple of circles. The present
knowledge on 2-knots, however,
does not allow us to say whether
this description makes any sense or should be discarded {\sl tout court}.
It can be taken for granted,
anyway, that uncovering the meaning of the amplitudes of this theory
is a source of insight for mathematics itself.

Finally, it is also interesting to know whether the properties
described so far
survive the coupling to matter. A positive answer will be given
in the next section, where explicit examples of the so-called
hyperinstantons introduced in ref.\ \cite{twist2,hyper,ghyp} by
Fr\`e and the
author
will be studied.

\subsection{Appendix: ${x_0\rightarrow \infty}$ does not contribute}
\label{appe}

Here I prove that the $x_0\rightarrow\infty$ $\partial{\cal M}$-component
does not contribute to the topological amplitudes.
The amplitude is always written as $\int_{\cal
M}\prod_{i=1}^n\omega_{\gamma_i}=
\int_{\partial {\cal M}}\Omega_{\gamma_1}\prod_{i=2}^n\omega_{\gamma_i}$
where $\omega_{\gamma_1}=d\Omega_{\gamma_1}$.
Since the $\gamma_i$'s are assumed to be compact, there exists an
$R$ such that
the 3-sphere of radius $R$ centered in the origin
contains any $\gamma_i$. Then, it is easy to prove, from (\ref{hal})
and (\ref{obser}), that
\begin{equation}
{\omega}_{\gamma_i}\sim {\rho^3R^{d_i}(\rho d^{4-d_i}x_0+x_0 d\rho\,
d^{3-d_i}x_0)\over (\rho^2+x_0^2)^4},
\quad
{\Omega}_{\gamma_i}\sim {R^{d_i}(3\rho^2+x_0^2)x_0d^{3-d_i}x_0\over
(\rho^2+x_0^2)^3},
\quad {\rm for}\,\, x_0\rightarrow \infty,
\end{equation}
where $d_i={\rm dim}\, \gamma_i$.
Due to the fact that $\sum_{i=1}^n{\rm codim}\, \gamma_i=5$, we have
$\sum_{i=1}^nd_i=4n-5$. Since $n\geq 2$, then
$\sum_{i=1}^nd_i\geq 3$. Let us assume $n=2$, which is the worst case.
Then, the $x_0\rightarrow \infty$ contribution is of the form
\begin{equation}
\int_0^\infty d\rho {R^3 (3\rho^2+x_0^2)x_0^5\rho^3\over
(\rho^2+x_0^2)^7}\sim
{R^3\over x_0^3}\rightarrow 0,
\end{equation}
as expected. Note that if we {\sl first} uncompactified some of the
$\gamma_i$'s ($R\rightarrow \infty$) and
{\sl then} took the limit $x_0\rightarrow \infty$ (which is
incorrect), we
would find problems.

\section{Matter coupling}
\label{coupling}
\setcounter{equation}{0}

In this section, the coupling to matter (scalar fields) is examined.
The ideas introduced in ref.s \cite{twist2,hyper,ghyp}
by Fr\'e and the author are used extensively.
Scalars can possess very
interesting instantons (called {\sl hyperinstantons}), that can be
coupled to
gravitational instantons \cite{twist2,hyper},
as well as Yang-Mills instantons \cite{ghyp}. Here we take
isospin $1/2$ scalar fields in the background of the Belavin et al.\
\cite{belavin}
instanton. The coupling constant $g$ is set to 1, for simplicity.
Hyperistantons are described by the lagrangian
(I convert to the Euclidean signature with respect to ref.\
\cite{ghyp})
\begin{equation}
{{\cal L}\over \sqrt{g}}=
{1\over 2} g^{\mu\nu}h_{ij}{\cal D}_\mu q^i{\cal D}_\nu q^j+
{1\over 4\beta} F_{\mu\nu}^a F^a_{\mu\nu}+
{\beta\over 2} {\cal P}^u_a{\cal P}^u_a.
\label{LK}
\end{equation}
A parameter $\beta$ has been introduced for future use.
This lagrangian is the bosonic piece of an N=2 lagrangian in a
special case
suggested by the topological twist
\cite{ghyp}. ${\cal L}$ can be written as
\begin{eqnarray}
{{\cal L}\over \sqrt{g}}&=&{1\over 8}
g^{\mu\nu}h_{ij}
\left({\cal D}_\mu q^i-\Lambda^{uv}{(j_u)_\mu}^\rho {\cal D}_\rho q^k
{(J_v)_k}^i\right)
\left({\cal D}_\nu q^j-\Lambda^{st}{(j_s)_\nu}^\sigma {\cal D}_\sigma q^l
{(J_t)_l}^j\right)\nonumber\\
&+&{1\over 2\beta}\left(F^{-a}_{\mu\nu}+
{\beta\over 2}\Lambda_{uv}I_{\mu\nu}^u{\cal P}^v_a\right)^2
+{1\over 8\beta}F^a F^a+{1\over 8}\Lambda_{uv}\Theta^u \hat \Omega^v.
\label{action}
\end{eqnarray}
$\Lambda^{uv}$ can be any $SO(3)$ matrix.
See \cite{ghyp} for the remaining notation, that will be in any case
explained along with the discussion. The instanton configurations are
\begin{equation}
F_{\mu\nu}^{-a}+{\beta\over 2}\Lambda_{uv}I_{\mu\nu}^u{\cal
P}^v_a=0,\quad\quad
{\cal D}_\mu q^i-\Lambda^{uv}
{(j_u)_\mu}^\nu {\cal D}_\nu q^j {(J_v)_j}^i=0.
\label{ki}
\end{equation}
I shall take $\beta\rightarrow 0$, so that the first equation
reduces to the
usual equation of Yang-Mills instantons, solved by the first of
(\ref{por}).
The other equation is invariant
and defines the hyperinstantons in the gauge-instanton background
(or gauged
triholomorphic maps, according to the mathematical interpretation
worked out in
\cite{hyper,ghyp}).
The topological properties should be independent of $\beta$, since
changing $\beta$ should be a continuous deformation of the
instanton equation.
It could happen, nevertheless, that some value of $\beta$ is
not reachable continuously, but I do not enter into these problems here.
After factorizing ${\rm exp}\left(-{1\over 8\beta}
F^aF^a\right)$ away,
 the limit $\beta\rightarrow 0$ in the Lagrangian ${\cal L}$
(\ref{action})
can be thought as a kind of Landau gauge
(indeed, from the point of view of topological field theory, the
instanton
conditions are simply
the gauge-fixing of the topological symmetry). Therefore, the $\beta=0$
solutions are {\sl hyperinstantons in the Landau gauge}.

To be explicit, in the case at hand $j_u=J_u=I^u=-\bar\eta_u$,
$\bar\eta_u$
denoting the anti-self-dual 't~Hooft symbols \cite{thooft},
$\Theta^u=dx^t I^u dx$, $\Omega^u=dq^t I^u dq$,
$\hat\Omega^u={\cal D}q^tI^u {\cal D}q+F^a{\cal P}^u_a$,
${\cal D}_\mu q^i=
\partial_\mu q^i+{1\over 2}A_\mu^a (\bar I_a)^{ij}q_j$,
${\cal P}^u_a={1\over 2}q^tI^u\bar I^a q$, $\bar I_a=-\eta_a$.
One has $d\Theta^u=d\Omega^u=d\hat \Omega^u=0$.
The index $i=1,\ldots 4$ goes over the real components of the isospin
$1/2$ representation.

Consider the following configuration
\begin{equation}
q^i(x)={(x-x_0)^i\over \rho \sqrt{\rho^2+(x-x_0)^2}}={(x-x_0)^i\over \rho
\sqrt{D}}.
\label{solut}
\end{equation}
One can check that ${\cal D}_\mu q^i={\delta_\mu^i\rho\over D^{3/2}}$.
The second equation of (\ref{ki}) gives
$\Lambda=\Lambda^t$ and ${\rm tr}\,\Lambda=-1$. We
choose $\Lambda={\rm diag}(1,-1,-1)$. In this way, the above scalar field
configuration,
which was known only as a solution to the equation $D_\mu D^\mu q^i=0$
\cite{thooft}, is also a hyperinstanton\footnotemark\footnotetext{As
a matter of fact, it is also a solution to certain vortex equations,
related to N=1 supersymmetry in the same way as the hyperinstanton
equations are related to N=2 supersymmetry.}.
This remark allows us to use the machinery developped in ref.\
\cite{twist2,hyper,ghyp}. Notice the crucial power $1/\rho$ in
(\ref{solut}).

Now, one wants to elaborate the specific observables of the above
scalar configuration and check whether they share the properties
of pure Yang-Mills theory.
The observables are related, via descent equations, to a
topological number,
called {\sl hyperinstanton number},
identified in \cite{hyper,ghyp} as the last term of (\ref{action}),
namely\footnotemark\footnotetext{As pointed out in \cite{ghyp}, eqs.\
(\ref{ki})
reduce to Witten's monopole equations \cite{monopoles} when $G=U(1)$ and
the manifold of the scalars $q$ is flat (in ref.s \cite{hyper,ghyp} the
$q$-manifold can be
a generic almost quaternionic manifold). The hyperinstanton number,
firstly introduced in \cite{hyper}, here plays a crucial role, but
it seems that in \cite{monopoles} there is no analogue of it.}
\begin{equation}
{\rm Hn}={1\over 4\pi^2}
\int_{M}\Lambda_{uv}\Theta^u\wedge \hat \Omega^v,
\label{hn}
\end{equation}
where $\hat \Omega^u=d\hat \omega^u$ and $\hat \omega^u=
q^t I_u dq+A^a {\cal P}_a^u$. One can check that Hn$=1$ for the solution
(\ref{solut}).
The instantons of the coupled theories are thus classified by {\sl
two} integer
numbers: the usual instanton number and the hyperinstanton number.

One could put an arbitrary constant $v$ in front of $q^i$, however such
a parameter should not be considered as a modulus:
the BRST variation of $v$ should be zero, otherwise
Hn$=v^2$ would not be
BRST invariant (see section 6 of ref.\ \cite{anomali} for an
analogous case
in topological abelian Yang-Mills theory coupled to topological gravity).
The eventual $v$-integration is made meaningful by the exponential factor
e$^{-{\rm Hn}}$. In practice, we can suppress the parameter $v$.

The desired observables are provided
by the descent equations associated to Hn in the usual
way \cite{baulieu,twist1,hyper}.
To begin with, let us consider
\begin{equation}
{\cal O}_{S^3_r}={1\over
4\pi^2}\int_{S^3_r}\Lambda_{uv}\Theta^u(2I^v_{ij}{\cal
D}q^i\xi^j+\psi_0^a{\cal P}^v_a)={1\over
4\pi^2}s\int_{S^3_r}\Lambda_{uv}\Theta^u\hat \omega^v=sf(r,\rho,x_0),
\end{equation}
where $s$ is the BRST operator (\ref{brstop})
and $\xi^i$ is defined by $sq^i=\xi^i-C_0^ak^i_a$,
$k^i_a(q)={1\over 2}(\bar
I_a)^{ij}q_j$. $\psi_0^a=\psi^a_{0\mu}dx^\mu$ and $C_0^a$ are given in
(\ref{por}).
One finds
\begin{equation}
f(r,\rho,x_0)={2\over \pi}\int_0^\pi{r^3 (r-x_0 \cos\theta)
\sin^2\theta\,
d\theta\over
(\rho^2+r^2+x_0^2-2 r x_0 \cos\theta)^2}.
\end{equation}
$f$ tends to $1$ for $r\rightarrow \infty$, i.e. to the
hyperinstanton number
Hn.
A generic amplitude ${\cal A}=<{\cal O}_{S^3_r}\cdot\prod_i {\cal
O}_{\gamma_i}>$
can be written as $\int_{\partial{\cal M}}f(r,\rho,x_0)\prod_i
\omega_{\gamma_i}$.
So, what matters is the limit of $f$ for $\rho\rightarrow 0$.
This turns out to be the familiar step function
\begin{equation}
\lim_{\rho\rightarrow 0}f(r,\rho,x_0)={1\over 2}(1+H(r-x_0)),
\end{equation}
precisely as in the case of pure Yang-Mills theory,
but now coming out of very different field configurations
(hyperinstantons) and observables related to a different
topological number (hyperinstanton number).

While the Yang-Mills instanton number generates ${\cal
M}$-differential forms
of any degree from 1 to 4 (compare with equations (\ref{hal}) and
(\ref{obser})
for $n=1$),
the hyperinstanton number  (\ref{hn}) generates only 1-forms
and 2-forms, obtained
by integrating its descendants over 3- and 2-dimensional
$M$-submanifolds,
respectively.
We have just  checked that in the first case the pure Yang-Mills
result is
reproduced.
To conclude,
we check the same thing in the second case, integrating over the 2-plane
$S^2=\{{\bf x}_2\}\otimes \IR^{\prime 2}$:
\begin{equation}
{\cal O}_{S^2}=\omega^{(2)}_{{\bf x}_2}=d\Omega^{(1)}_{{\bf
x}_2},\quad\quad\quad
\Omega^{(1)}_{{\bf x}_2}={1\over
4\pi^2}\int_{S^2}\Lambda_{uv}\Theta^u\omega^v_{(0,1)},
\end{equation}
$\omega^u_{(0,1)}$ being the first descendant of $\hat \omega^u$.
One finds
\begin{equation}
\xi^i=-{\rho dx_0^i\over D^{3/2}}
-{(x-x_0)^i[2\rho^2+(x-x_0)^2]d\rho\over \rho^2 D^{3/2}},\quad\quad
\omega^u_{(0,1)}=q^tI_u\xi=-{(x-x_0)^tI^udx_0\over D^2}.
\end{equation}
Using the $I^u$-basis of ref.\ \cite{ghyp}, one has
\begin{equation}
\Omega^{(1)}_{{\bf x}_2}={1\over
2\pi}{(x^0_2-x^0_0)dx_0^1-(x^1_2-x^1_0)dx_0^0\over \rho^2+({\bf x}_2-{\bf
x}_0)^2}.
\end{equation}
The pure Yang-Mills analogue of this expression is written in
formula (4.41) of
ref.\ \cite{anomali}. The two expressions are indeed different.
However, when
inserted into an amplitude,
for example, $<{\cal O}_{S^2}\cdot {\cal O}_{S^1}>=\backslash\!\!\!\slash
(S^2,S^1)$,
only the $\rho\rightarrow 0$ limit matters  and both expressions
have the same
$\rho\rightarrow 0$ limit, apart from an immaterial sign and the
normalization
(the overal
factor ${1\over 16\pi^2}$, not introduced in \cite{anomali} from
the begining).
So, again, the coupling to matter agrees with the pure theory,
normalization
included.

\section{Topological embedding}
\label{embedding}
\setcounter{equation}{0}

The purpose of this section and the next one
is to study the relevance of link invariants and topological field
theory for
physics.
In the end of the day the properties found in ref.\ \cite{anomali} and in
the previous sections are
properties of very special solutions to the field equations of pure QCD.
It is hard to believe that such properties have no relation with physics.

The first aim is to embed  topological Yang-Mills theory
into ordinary Yang-Mills theory and show in what limit (and for what
amplitudes)
the latter reduces to the former. In the next section the match
with physical
phenomena is
discussed.

The idea is that the topological version of a theory is a useful
device for
defining perturbation theory in the topologically nontrivial sectors of
the same theory\footnotemark\footnotetext{The `classical'
approach to this problem can be found, for example in
\cite{rajaraman,coleman}.
The approach followed here is different under several aspects, but
contains
the old one.}.
It allows one to separate in a convenient way
the nonperturbative part, which is the integration over the
instanton moduli space ${\cal M}$, from the perturbative part, described
by the quantum
fluctuations around the instanton `vacuum'. At the same time, the
topological theory can be recovered as a limit of the ordinary
theory, when the quantum fluctuations are suppressed.
The embedding of the former into the latter,
which will be called {\sl topological embedding},
is just a generalization
of the usual procedure of treating the collective coordinates by
``introducing
1'' {\sl a la Faddeev-Popov} \cite{faddeev,patrasciou}: here
I suggest to treat the topologically nontrivial
sectors of a theory by ``introducing
the topological
version'' of the same theory.
The observables of the topological theory
are useful to define the (previously ill-defined) integration over
collective
coordinates.
Of course, there
are many inequivalent choices. One of them,
the insertion of the volume
form of the instanton moduli space ${\cal M}$,
gives back the common result, in which the
(infinite) volume of the moduli space
appears as an overall factor.
The other possibilities
offered by the topological embedding and
not contemplated within the usual approach,
are indeed the insertions ${\cal M}$-top-forms
made by products of
topological observables ${\cal O}_{\gamma_i}$.
In this approach, the infinite volume factor gets ``regularized'' and
perturbation theory in the
topologically nontrivial sectors is consistent.

In view of this, the amplitudes of the
topological version of the theory give
information about the nonperturbative
nature of the complete theory. They are a useful device
for extracting otherwise invisible properties
of the instantonic configurations, that should have some physical
meaning,
if instantons do.
Applying the results of ref.\ \cite{anomali} and the previous
sections to QCD,
it will be argued that the step amplitudes
computed there and generalized
here are related to a non abelian generalization of the
Aharonov-Bohm effect.

To be explicit,
I focus on topological Yang-Mills theory with $G=SU(2)$, $M=\IR^4$
in the $k=1$
sector, in which case we have all the
needed explicit formul\ae. Let us decompose the gauge connection $A$
into $A_0+gA_q$, $A_0$ being the Belavin et al.\ instanton
\cite{belavin} and
$A_q$ denoting the quantum fluctuation around it. Now,  $A_0$ spans
a moduli
space ${\cal M}$, while
$A_q$ is restricted to be perperdicular to it, otherwise
tangential fluctuations are counted twice: once in $A_0$ and a
second time in $A_q$.
Having done this decomposition, we know that the integral over $A_q$ is
well-defined. For example, in the quadratic approximantion, it
gives the primed
determinant of the kinetic operator, which has been computed
explicitly by 't
Hooft
in ref.\ \cite{thooft}. The ${\cal M}$-integral, on the other hand,
is not
well-defined. Usually
\cite{coleman,rajaraman}, one takes the attitude that the problem
is due to
perturbation theory and should disappear in the exact answer. This
is not so
useful from the practical point of view
and is equivalent to declare that the perturbative expansion around
instantons
is inconsistent.
Here I take a radically different attitude. The key idea is to say
that there
is a gauge-symmetry that has not been gauge-fixed before and that the
ill-definition of the ${\cal M}$-integral is
like the ill-definition of any functional integral before fixing ordinary
geuge-symmetry.
Indeed, there is no way, from the physical point of view, to
privilege any
$m\in {\cal M}$
(any position or size of the instanton, for example), because any
$m$ is a
minimum of the classical action. So, in the context of the perturbation
expansion that I want to consider,
deforming $m$ around ${\cal M}$ is a gauge-symmetry. It is exactly a
topological symmetry: the most general continuous deformation of
the instanton
in the space of instantons.
To cure the problem, we generalize the usual BRS recipe to global
degrees of
freedom.
We have to do two things: i) to introduce ghosts (to be called {\sl
topological
ghosts})
associated with the topological symmetry
while preserving the nilpotence of the BRS operator; ii) to
introduce a BRS
closed
operator that gauge-fixes the symmetry, making both the ${\cal
M}$-integral and
the integral
over the topological ghosts meaningful. These aims can be achieved
by embedding
the topological
version of the theory into the physical theory.

The topological ghosts are just the ghosts of topological
Yang-Mills theory.
So, let us write the BRST algebra of ordinary Yang-Mills theory
\begin{equation}
sA^a=-DC^a,\quad\quad\quad sC^a=-{1\over 2}\varepsilon_{abc}C^bC^c,
\label{sum}
\end{equation}
as the semidirect product of the BRST algebra of topological
Yang-Mills theory
\begin{eqnarray}
sA_{0\mu}^a&=&\psi_{0\mu}^a+D_\mu(A_0) C_0^a,\quad\quad
s\psi_{0\mu}^a=-D_\mu\phi_0^a-{\varepsilon^a}_{bc}\psi^b_{0\mu}
C_0^c,\nonumber\\
s\phi_0^a&=&{\varepsilon^a}_{bc}\phi_0^bC_0^c,\quad\quad\quad
\quad\quad
sC_0^a=\phi_0^a-{1\over 2}{\varepsilon ^a}_{bc}C_0^bC_0^c,
\label{brsym}
\end{eqnarray}
times the following ``remnant''
\begin{eqnarray}
sA_{q\mu}^a&=&-{1\over g}\psi_{0\mu}^a+D_\mu (A)C_q^a+{\varepsilon
^a}_{bc}A_{q\mu}^bC_0^c,\nonumber\\
sC_q^a&=&-{1\over g}\phi_0^a
-{\varepsilon ^a}_{bc}C_0^bC_q^c-{g\over 2}{\varepsilon
^a}_{bc}C_q^bC_q^c.
\label{porb}
\end{eqnarray}
(\ref{sum}) is the sum of  (\ref{brsym}) and (\ref{porb}), $C^a$ being
identified with $C^a_0+gC^a_q$. The complete functional integral is
obtained by
integrating over the  `topological fields' $A_0$, $\psi_0$, $C_0$
and $\phi_0$
and the quantum fluctuations $A_q$ and $C_q$.
The functional measure over the topological fields reduces to
$\int_{\cal M}dm \,d\hat m$, $m$ denoting the moduli
and $\hat m$ being their ghost partners ($\hat m=sm$).
The topological subset (\ref{brsym}) is closed under
BRS transformations. This fact, as we know from \cite{anomali},
allows us to
``solve the BRS algebra'', i.e.\ find the explicit expressions of the
topological ghosts  that follow from nilpotence and the explicit
expression of
$A_0$.
The complete solution is
\cite{anomali}
\begin{eqnarray}
A_0^a&=&{2\over D}\,
dx^\mu\eta^a_{\mu\nu}(x-x_0)^\nu,\quad\quad\quad\quad\quad
C_0^a=-{2\over D}\, \hat x_0^\mu\eta^a_{\mu\nu}(x-x_0)^\nu,\nonumber\\
\psi^a_{0\mu}&=&-4 {\rho\over D^2}\eta^a_{\mu \nu}[ \rho \hat
x_0^\nu+(x-x_0)^\nu \hat\rho],
\quad
\phi_0^a=-{2\rho\over D^2}\, \eta^a_{\mu\nu}[\rho \hat
x_0^\mu+2(x-x_0)^\mu
\hat\rho)]\hat x_0^\nu.
\label{por}
\end{eqnarray}
It is worth recalling that $\psi_{0}$ is gauge-fixed with the condition
$D_\mu(A_0)\psi_{0\mu}^a=0$.
Finally, we can write the generating functional $Z[J_q]$  as
\begin{equation}
Z[J_q]=\sum_{k\in\IZ}{\rm e}^{i\theta k-{8\pi\over
g^2}|k|}\int_{{\cal M}_k}dm
\,Z_k[m,J_q],
\end{equation}
where $Z_k[m,J_q]$ is the partition function for a fixed value
of the instanton moduli $m$ with
instanton number $k$, namely
\begin{equation}
Z_k[m,J_q]=\int d\hat m\int dA_q dC_q d\mu_{gf}\,\,{\rm exp}\left[-{\cal
S}(A_q,m,\hat m)+{\cal S}_{gf}(A_q,m,\hat m)+J_qA_q
\right].
\label{right}
\end{equation}
${\cal S}_{gf}(A_q,m,\hat m)$ denotes a suitable gauge-fixing term and
$d\mu_{gf}$ is the relevant functional integration measure for Lagrange
multipliers and antighosts.
In this way, we have conveniently separated the nonperturbative and the
perturbative aspects of the topologically nontrivial sectors
of Yang-Mills theory.
$Z[m,J_q]$ can be calculated perturbatively, since the background
is fixed.
The BRST algebra is (\ref{porb}) and the quantum fluctuations $A_q$ are
restricted to be perpendicular to ${\cal M}$. They
have a well-defined propagator and one can safely define Feynmann rules.
The perturbative amplitude is in general $m$-dependent and
the final amplitude is obtained after the (non-perturbative)
integration over
${\cal M}$.
The amplitudes computed in \cite{anomali} and in the previous section
are examples in which the full contribution comes from the
nonperturbative
part.
Nevertheless, it is clear that they are also very peculiar amplitudes of
ordinary Yang-Mills theory in the topologically nontrivial sectors.

The action ${\cal S}(A_q,m,\hat m)$ is obtained by expanding
the usual Yang-Mills action around $A_0$:
\begin{eqnarray}
{\cal S}(A_q,m,\hat m)&=&{1\over
4}[D_\mu(A_0)A^a_{q\nu}-D_\nu(A_0)A^a_{q\mu}]^2
+{1\over 2}\varepsilon_{abc}F^a_{\mu\nu}(A_0)A^b_{q\mu}A^c_{q\nu}
\nonumber\\
&+&g\varepsilon_{abc}A_{q\mu}^aA_{q\nu}^b
D_\mu(A_0)A^c_{q\nu}+{g^2\over 4}
(\varepsilon_{abc}A_{q\mu}^bA_{q\nu}^c)^2.
\end{eqnarray}
The gauge-fixing term is made by two pieces:
the first term fixes
the gauge symmetry $\delta A_{q\mu}^a=D_\mu C_q^a$, while the second one
fixes the topological symmetry. The first gauge-fixing is achieved
with the
usual
condition $D_\mu(A_0)A_{q\mu}^a=0$, thus preserving the topological
symmetry.
So, the ordinary gauge-fixing term is
\begin{equation}
{\cal S}_{gf}(A_q,m,\hat m)={1\over 2}(D_\mu(A_0)A_{q\mu}^a)^2+\bar
C_q^a[D_\mu(A_0)D_\mu(A)C_q^a+\varepsilon_{abc}\psi_{0\mu}^bA_{q\mu}^c+
\varepsilon_{abc}D_\mu(A_0)A_{q\mu}^bC_0^c].
\end{equation}
The quadratic part ${\cal S}_Q$
of the gauge-fixed action ${\cal S}+{\cal S}_{gf}$ mixes $A_q$ and
the ghosts
$\bar C_q$ and $C_q$ in a nontrivial way: defining
$V_q=[A_q,C_q,\bar C_q]$ and
$Q(m,\hat m)$ such that  ${\cal S}_Q=V_q^t Q(m,\hat m) V_q$, the
zero modes are
collected in the vector $[\psi_0,\phi_0,0]$: $Q(m,\hat
m)[\psi_0,\phi_0,0]=0$.

Finally, let us discuss point ii) mentioned above, namely how to
gauge-fix
the topological symmetry.
Usually, to gauge-fix a symmetry one introduces and operator of the
form ${\cal
O}=
\delta({\cal G})\, s{\cal G}$, ${\cal G}$ being the chosen gauge-fixing
condition.
${\cal O}$ is clearly BRS closed and makes both the integration over the
gauge-fields
(via $\delta({\cal G})$) and the integration over the gauge-ghosts
(via $s{\cal
G}$)
meaningful. The moral of the story is that we have to introduce a
BRS-closed
operator
${\cal O}$ that makes both the $m$- and $\hat m$-integrations meaningful.
It is now clear that ${\cal O}$ has to be constructed with
the topological observables ${\cal O}_{\gamma_i}$, i.e.\
${\cal O}=\prod_i {\cal O}_{\gamma_i}$. Concretely,
one can  modify (\ref{right}) into
\begin{equation}
Z_k[m,J_q,\zeta]=\int d\hat m\int dA_q dC_q d\mu_{gf}\,\,{\rm
exp}\left[-{\cal
S}(A_q,m,\hat m)+{\cal S}_{gf}(A_q,m,\hat m)+J_qA_q+\zeta(\gamma){\cal
O}_{\gamma}
\right],
\label{right2}
\end{equation}
the term $\zeta(\gamma){\cal O}_{\gamma}$ standing for all the possible
insertions of topological observables.
't Hooft's choice \cite{thooft}, instead, dictated by a simple
dimensional
argument, is
\begin{equation}
{\cal O}={\hat \rho\prod_{\mu=1}^4(\hat x_0)_\mu\over \rho^5}.
\label{too}
\end{equation}
It does not fix the $x_0$-translations and leave the problem of the
$\rho$-integration open.

Now, let us take a certain number of $\zeta$-functional derivatives
in order to
introduce a product of ${\cal O}_{\gamma_i}$ that saturates the
moduli space
dimension. Due to this, any term in ${\cal S}_{gf}$ containing
$\psi_0$ or
$C_0$
can be dropped and the action ${\cal S}+{\cal S}_{gf}$ reduces to
the usual
one.
In the $g\rightarrow 0$ limit, only the quadratic part matters, which
integrates to the primed determinants that combine with the ${\rm
e}^{-{8\pi\over g^2}}$ factor
to give a renormalization group invariant expression \cite{thooft}.
If we focus, for now, on dimensionless amplitudes (in some sense, the
`partition functions' of the topologically nontrivial sectors), i.e.\
amplitudes with no gluons $A_q$, this expression is just a
constant\footnotemark\footnotetext{For more detailed explanations
on how this
happens and its
implications, see section 2.1 of ref.\ \cite{scond}.}.
Finally, the ${\cal M}$ integration gives back the topological amplitude
associated to the ${\cal O}_{\gamma_i}$'s, namely
\begin{equation}
{\rm const.}\,{\rm e}^{i\theta}\int_{\cal M}dm\, d\hat m \,
\prod^n_{i}{\cal
O}_{\gamma_i}(m,\hat m).
\label{choice}
\end{equation}
A choice like (\ref{choice}) `renormalizes' the infinite factor
that would be
obtained with the 't Hooft measure (\ref{too}).
The (physically meaningful)
freedom related to the choice of (\ref{choice}) can also be thought
as the arbitrariness associated to the renormalization of the
infinite volume
factor.
When there are gluons in the amplitude, (\ref{choice}) is just the
measure over
the moduli space.
The amplitude has the form
\begin{equation}
{\rm const.}\,{\rm e}^{i\theta}\int_{\cal M}dm\, d\hat m \,
\prod^n_{i}{\cal
O}_{\gamma_i}(m,\hat m)\,{\cal A}(x_1,\ldots x_n;m),
\end{equation}
${\cal A}(x_1,\ldots x_n;m)$ denoting the perturbative amplitude at
fixed $m$.
It is clear that the measure (\ref{choice}) makes the ${\cal M}$-integral
convergent.
We conclude that the physical meaning of topological field theory
is that it
provides
the set of consistent measures over the moduli space.

Due to the nature of the topological symmetry and its gauge-fixing,
it is clear that the theory ``depends'' on the gauge-fixing itself,
namely on
the choice os $\gamma_i$'s,
in the sense that two gauge-fixings that are not continuously related
to each other give different answers. It is very nontrivial, in general,
to solve the problem of classifying the gauge-fixing dependence of
a theory.
Nevertheless,
for the specific aspect of the problem that we are considering now,
namely the
dependence on the ``gauge-fixing for collective coordinates'' in the unit
instanton number sector of $SU(2)$ (or $SU(N)$) Yang-Mills theory,
we already know the answer, that is the multilink intersection theory
elaborated n the previous sections.
To be explicit, the unlinking process described in section
\ref{multilinks}
is a non-continuous deformation whenever a complete
multilink intersection is crossed.
It is a continuous deformation for any incomplete intersection.
That is why sometimes I speak about topological ``gauge-fixing''
and other
times
about ``insertion of  topological observables'',
when referring to this aspect of the theory. In other words,
{\sl such a gauge-fixing choice is observable}. Actually, our
approach allows
us to say that
the topological aspects are the only genuine instantonic properties
that can be observable. Dynamics only comes form the quantum fluctuations
$A_q$,
i.e.\ gluons propagating over the instanton background specified by
$\prod_i{\cal O}_{\gamma_i}$.

A comment is in order about the regularization technique that is most
convenient
to treat perturbation theory around instantons (see also \cite{thooft}).
The dimensional technique presents some problems, since instantons
are purely
four dimensional objects.
A Pauli-Villars-type technique seems to be better.
To avoid the problem that the mass terms of the regulators
break gauge-invariance, one can use the following alternative
regularization
technique,
firstly defined and
used by Johansen in \cite{johansen} within the usual perturbation theory
framework.
Embed Yang-Mills theory into N=4 super Yang-Mills theory and break N=4
supersymmetry down to N=0 by giving mass terms to each additional field.
For finite masses, the theory is finite and so is a good regulator.
When one
wants to recover the
initial theory, one has to let the masses tend to infinity and
the divergent terms have then
to be subtracted with the usual renormalization
algorithms.

To conclude, topological Yang-Mills theory is a useful device to define
perturbation theory in the topologically nontrivial sectors
of ordinary Yang-Mills theory and is also
a certain limit of this theory itself. The topological amplidutes
calculated so
far
can play a role in the ``real theory''.

\section{Non abelian analogue of the Aharonov-Bohm effect}
\label{aharonov}
\setcounter{equation}{0}

In this section, the relation between link invariants and physics
is discussed,
focusing, in particular on a non abelian analogue of the
Aharonov-Bohm effect.

QCD is expected to confine and confinement is nonperturbative in nature.
However, it is very difficult to get
nonperturbative information about Yang-Mills theory.
There are some nonperturbative
aspects, nevertheless, (like instantons) that can be studied exactly,
for example using the topological field theoretical device.
These aspects of the theory could
carry some, perhaps unconventional, sign
that the theory confines and
the link invariants could be interpreted as such a sign. The first
thing that
comes to one's mind
when speaking about link invariants is the Aharonov-Bohm effect.

In Maxwell theory, a wire $\gamma$ of magnetic dipoles with
magnetic moment $g$
per unit length (a thin circular solenoid, for example) generates a
vector potential ${\bf A}({\bf x},\gamma)$ equal to
\begin{equation}
{\bf A}({\bf x}^\prime,\gamma)=-g\oint_\gamma d{\bf x}\wedge \nabla
{1\over |{\bf x}-{\bf x}^\prime|},
\end{equation}
so that the magnetic flux associated to a loop $\gamma^\prime$ is
given by
Gauss' formula
(\ref{gauss}):
\begin{equation}
\Phi_{\gamma^\prime}({\bf B}_{\gamma})=\oint_{\gamma^\prime}
{\bf A}({\bf x}^\prime,\gamma)\cdot d{\bf x}^\prime=4\pi g \,\, \gamma
\backslash\!\!\!\slash \gamma^\prime.
\end{equation}
The flux is quantized, according to the link number of the two loops.
This  is physically meaningful, since it can be observed (Aharonov-Bohm
effect).
The noticeable property of the field configuration is that both the
electric
and magnetic fields are zero everywhere except than within
$\gamma$. So, in
$\gamma$ there is a concentration
of three entities: the gauge field, the positively charged matter and the
negatively charged matter.

Quantum nonabelian gauge field theory should have the property that
three (or more) entities like the above ones (i.e.\ the gauge bosons)
are forced by the gauge symmetry to always screen one another.
At least, this should be visible in any exact amplitude.
Take, for example, $SU(2)$ gauge field theory
and look at it as a $U(1)$ gauge theory (previliging some vector
boson, which
we call the photon) coupled to charged matter (the other two gauge
bosons).
Due to the underlying gauge symmetry, the ``matter''
and the ``photon'' are mixed together and indistinguishable.
We know that link numbers appear in Maxwell theory
plus matter only when
the field strength and the charged matter are confined together.
We also know, from the computations made in this paper, that
link numbers in nonabelian gauge field theory
appear naturally in special, but exact, amplitudes of the theory.
That means precisely that,
in the realm of the amplitudes here computed,
which are the only exact nonperturbative amplitudes available so far,
QCD, when regarded as a matter coupled QED,
is such that the ``matter'' naturally screens the photon.
Due to the underlying gauge symmetry, this holds for any ``photon'' and
consequently for
the entire group. In this sense, the link numbers that we have
found can be
regarded as
an unconventional sign of confinement.
A nontrivial consequence of this description is that, although
QCD confines,  the non-abelian analogue of the Aharonov-Bohm
effect cannot be ``screened'' and
should be experimentally observable. Maybe in some future we shall
be able to
construct wires
carrying a nonvanishing color current and
create something similar to the magnetic force lines that penetrate the
superconductors
\cite{scond}. Something like this should correspond to making
experiments in a
sector
specified by (\ref{choice}).

Since confinement is expected to be a nonperturbative phenomenon,
one should resum the perturbative series before being able to reveal it.
In the perturbative framework, confinement is hardly visible,
since when the coupling constant is zero the theory becomes
practically abelian and an abelian theory does not confine.
So, it  may be very interesting
to investigate some genuinely
nonperturbative aspects of the theory like the ones considered in
this paper.
The amplitudes computed here represent special situations in which one is
allowed to freeze the quantum fluctuations (so that no resummation
at all is
necessary)
and yet get something nontrivial. Finally, it is clear that
the topological embedding is an
approach that also deserves to be studied {\sl per se}.

Link numbers in QCD appear in the pure theory, i.e.\ they do not
require the
presence of matter.
In section \ref{coupling}, we saw that they can also appear in
presence of
matter.

\section{Conclusions}
\label{concl}
\setcounter{equation}{0}

In ref.\ \cite{anomali} and the present paper, new properties of the the
Belavin-Polyakov-Schwarz-Tyupkin
instanton were uncovered, using topological field theory as a tool.
An unexpected connection with link theory came out.
In \cite{anomali} the presence of {\sl some} link theory was
detected, here
the theory was identified completely.
The feeling is that higher instanton numbers \cite{rebbi,atiyah}
hide deeper mathematical concepts.
No trace of vanishing amplitudes has been found so far, so one open
problem
(of the many) is to identify the mathematical meaning of the
amplitudes with
higher instanton number.

The second point concerns physics. The physical role of
topological field
theory
was clarified by showing that its embedding in the associated
physical theory
is useful to perform the perturbative expansion in the
topologically nontrivial
sectors.
The topological properties of  very
special solutions to the Yang-Mills field equations, like instantons are,
are not expected to be unrelated to physics. Here they are
related to a non abelian version of the Aharonov-Bohm effect.

\begin{center}
{\bf Acknowledgements}
\end{center}

I would like to thank V.\ Schomerus for many interesting discussions
on 2-knots, link theory and related subjects.
I also thank A.\ Johansen for discussions.
This research was supported in part by the
Packard Foundation and by NSF grant PHY-92-18167.


\vskip .3in


\begin{thebibliography}{99}

\bibitem{rossi} D.\ Amati, K.\ Konishi, Y.\ Meurice, G.C.\ Rossi and G.\
Veneziano,
Non-perturbative aspects in supersymmetric gauge-theories,
Phys.\ Rep.\ 162 (1988) 169.

\bibitem{wittenym}
E.\ Witten, Topological quantum field theories,
Comm. Math. Phys. 117 (1988) 353.

\bibitem{twist1} D.\ Anselmi and P. Fr\`e, Twisted N=2 supergravity as
topological gravity in four dimensions, Nucl.\ Phys.\ B392 (1993) 401.

\bibitem{twist2} D.\ Anselmi and P.\ Fr\`e, Topological twist in four
dimensions, R-duality and hyperinstantons, Nucl.\ Phys.\ B404 (1993) 288.

\bibitem{martellini} A.S.\ Cattaneo, P.\ Cotta-Ramusino, A.\ Gamba
and M.\
Martellini,
The Donaldson-Witten invariants in pure 4D-QCD with order and disorder 't
Hooft-like operators,
preprint IFUM 493/FT and hepth/9502110.

\bibitem{donaldson}
S.K. Donaldson, An application of gauge theories to the topology of four
manifolds,
J. Diff. Geom. 18 (1983) 269.

\bibitem{anomali} D.\ Anselmi, Anomalies in instanton calculus,
Nucl.\ Phys.\ B439 (1995) 617.

\bibitem{hyper} D.\ Anselmi and P. Fr\`e, Topological sigma models
in four
dimensions and triholomorphic maps, Nucl.\ Phys.\ B416 (1994) 255.

\bibitem{ghyp} D.\ Anselmi and P. Fr\`e, Gauged hyperinstantons and
monopole
equations,
Phys.\ Lett.\ 347B (1994) 247.

\bibitem{baulieu}
L.~Baulieu and I.M. Singer, Topological Yang-Mills theory, Nucl.
Phys. B (proc.
suppl.) 5 B (1988) 12.

\bibitem{rebbi} R.\ Jackiw, C.\ Nohl and C.\ Rebbi, Conformal
properties of
pseudoparticle configurations, Phys.\ Rev.\ D 15 (1977) 1642.

\bibitem{atiyah} M.F.\ Atiyah, N.J.\ Hitchin, V.G.\ Drinfeld and
Yu.I.\ Manin,
Construction of instantons, Phys.\ Lett.\ 65A (1978) 185.

\bibitem{belavin} A.A.\ Belavin, A.M.\ Polyakov, A.S.\ Schwarz and Yu.S.\
Tyupkin,
Pseudoparticle solutions of the Yang-Mills equations, Phys.\ Lett.\
59B (1975)
85.

\bibitem{thooft} G.\  't~Hooft, Computation of the quantum effects
due to a
four-dimensional
pseudoparticle, Phys.\ Rev.\ D 14 (1976) 3432.

\bibitem{monopoles} E.\ Witten, Monopoles and four manifolds,
Math.\ Research
Lett.\ 1 (1994) 769.

\bibitem{rajaraman} R.\ Rajaraman, Solitons and Instantons,
North Holland, 1982.

\bibitem{coleman} S.\ Coleman, Aspects of symmetry, Cambridge
University Press,
1985.

\bibitem{faddeev} L.D.\ Faddeev and V.N.\ Popov, Phys.\ Lett.\ 25B
(1967) 29.

\bibitem{patrasciou} E.\ Gildener and A.\ Patrasciou,
Pseudoparticle contributions to the energy spectrum of a one-dimensional
system,
Phys.\ Rev.\ D16 (1977) 423.

\bibitem{scond} D.\ Anselmi, On field theory quantization around
instantons,
preprint HUTP-95/A026, July 195.

\bibitem{johansen} A.A.\ Johansen, Ultraviolet regularization of N=1
supersymmetric theories by means of extended supersymmetries, Sov.\
J.\ Nucl.\
Phys.\ 45 (1987) 167.

\end{thebibliography}
\end{document}